\documentclass[twocolumn,aps,prd,superscriptaddress]{revtex4-1}  
\usepackage{amsmath,graphicx}

\DeclareMathOperator{\Tr}{tr}

\usepackage{natbib}
\usepackage{amsmath,amssymb}

\usepackage[usenames]{color}

\newcommand{\Agw}{\ensuremath{A_\mathrm{gw}}}
\newcommand{\res}{\ensuremath{\delta {\bf t}}}

\begin{document}

\title{Noise-marginalized optimal statistic: A robust hybrid frequentist-Bayesian statistic for the stochastic gravitational-wave background in pulsar timing arrays}

\author{Sarah J.\ Vigeland}
\affiliation{Center for Gravitation, Cosmology and Astrophysics, University of Wisconsin--Milwaukee, PO Box 413, Milwaukee WI, 53201, USA}

\author{Kristina Islo}
\affiliation{Center for Gravitation, Cosmology and Astrophysics, University of Wisconsin--Milwaukee, PO Box 413, Milwaukee WI, 53201, USA}

\author{Stephen R.\ Taylor}
\affiliation{Theoretical AstroPhysics Including Relativity (TAPIR), MC 350-17, California Institute of Technology, Pasadena, California 91125, USA}
\affiliation{Jet Propulsion Laboratory, California Institute of Technology, 4800 Oak Grove Drive, Pasadena, California 91106, USA}

\author{Justin A.\ Ellis}
\affiliation{Department of Physics and Astronomy, West Virginia University, P.O. Box 6315, Morgantown, WV 26506, USA}
\affiliation{Center for Gravitational Waves and Cosmology, West Virginia University, Chestnut Ridge Research Building, Morgantown, WV 26505, USA}
\affiliation{Infinia ML, 202 Rigsbee Avenue, Durham, NC 27701, USA}


\date{\today}  

\begin{abstract}
Observations have revealed that nearly all galaxies 
contain supermassive black holes (SMBHs) at their centers. 
When galaxies merge, these SMBHs form SMBH binaries (SMBHBs) 
that emit low-frequency gravitational waves (GWs). 
The incoherent superposition of these sources produce a stochastic GW background (GWB) 
that can be observed by pulsar timing arrays (PTAs). 
The optimal statistic is a frequentist estimator of the amplitude of the GWB 
that specifically looks for the spatial correlations between pulsars induced by the GWB. 
In this paper, we introduce an improved method for computing the optimal statistic 
that marginalizes over the red noise in individual pulsars. 
We use simulations to demonstrate that this method 
more accurately determines the strength of the GWB, 
and we use the noise-marginalized optimal statistic 
to compare the significance of monopole, dipole, and Hellings-Downs (HD) spatial correlations 
and perform sky scrambles.
\end{abstract}

\maketitle

\section{Introduction}

Long-wavelength gravitational waves (GWs) with frequencies of 
$10^{-9} - 10^{-7} \; \mathrm{Hz}$ can be observed with pulsar timing arrays (PTAs) 
composed of millisecond pulsars (MSPs) \cite{hd1983,fb1990}. 
The dominant astrophysical source in this frequency range is the isotropic stochastic 
gravitational wave background (GWB) 
made up of the incoherent superposition of GWs from inspiraling 
supermassive black hole binaries (SMBHBs) 
\citep{1995ApJ...446..543R, 2003ApJ...583..616J, 2003ApJ...590..691W}. 
By monitoring the periodic emission from these pulsars using radio telescopes, 
we can probe the dynamics of the spacetime through which the pulses travel. 
This is done by searching for correlations in the pulsar timing residuals, 
which measure the differences between the expected and observed pulse times of arrival (TOAs). 
Current upper limits on the stochastic background from PTAs are approaching 
theoretical predictions for the GWB \citep{PPTA2013,EPTA2015,abb+17b}

PTAs primarily use Bayesian data analysis to compare the inferred probabilities 
of various models for the residuals, 
including one where they contain the GWB 
\citep{vlm+2009,lah+2013}. 
Bayesian inference is a powerful tool because 
it properly accounts for degeneracies between parameters 
and incorporates all sources of uncertainty into the analysis. 
However, running a full Bayesian analysis is computationally intensive, 
particularly when searching for evidence of Hellings-Downs (HD) spatial correlations --
the ``smoking gun'' of the GWB.

The significance of the GWB can also be assessed using the 
optimal statistic, a frequentist estimator for the GWB amplitude \citep{abc+2009,demorest+2013,ccs+2015}. 
Not only does it provide an independent detection procedure, complementing a more robust Bayesian analysis, but it requires significantly less time to compute. In particular, the optimal statistic produces results for a given spatial correlation function within seconds; a full Bayesian analysis including correlations has to run for many weeks on a supercomputing cluster. 

However, when pulsars have significant red noise 
the optimal statistic gives biased results 
due to the strong covariance 
between the individual red noise parameters and the GWB amplitude. 
Many individual pulsars show evidence for red noise \citep{lam+2017,abb+17}, 
and uncertainty in the position of the Solar System barycenter (SSB) leads to 
a common red process in all pulsars \citep{abb+17b}. 
Here we present a technique for improving the accuracy of the optimal statistic by including an additional step: 
marginalizing over the individual pulsars' red noise parameters 
using the posterior distributions from a full Bayesian analysis of all the pulsars. 
This hybrid approach produces a more precise estimate of the GWB amplitude \Agw\ 
and its uncertainty, while requiring only a few minutes more than the more traditional method of computation. 
Furthermore, the same Bayesian analysis drawn upon by the noise marginalization 
can be used to compute the optimal statistic for any choice of spatial correlations 
simply by changing the overlap-reduction function (ORF). 
For example, clock errors lead to a common red signal with monopole spatial correlations \citep{hcm+2012}, 
while uncertainty in the SSB produces dipole spatial correlations \citep{chm+2010}. 
This technique is used to perform the frequentist searches 
for common red signals with HD, 
monopole, and dipole spatial correlations 
in the NANOGrav 11-year data set \citep{abb+17b}.

This paper is organized as follows. In Sec.~\ref{sec:marg_os} 
we lay out the procedure for computing the noise-marginalized optimal statistic. 
We use simulations to compare the noise-marginalized optimal statistic 
to the optimal statistic computed with fixed noise. 
In Sec.~\ref{sec:spatial} we determine how well 
the noise-marginalized optimal statistic can 
differentiate between monopole, dipole, and HD spatial correlations. 
In Sec.~\ref{sec:skyscrambles} we use the noise-marginalized optimal statistic 
to perform sky scrambles, 
which assess the significance of HD spatial correlations 
by scrambling the pulsars' sky positions \citep{cs2016, tlb+2017}. 
We summarize our results in Sec.~\ref{sec:conclusion} 
as well as discuss future applications of the noise-marginalized optimal statistic.

\section{Noise-marginalized optimal statistic}
\label{sec:marg_os}

The optimal statistic is a frequentist estimator 
for the amplitude of an isotropic stochastic GWB, 
and can be derived by analytically maximizing 
the PTA likelihood function in the weak-signal regime \citep{abc+2009,ccs+2015}. 
It is constructed from the timing residuals $\res$, 
which can be written as 
\begin{equation}
	\res = M {\bf{\epsilon}} + F {\bf{a}} + U {\bf{j}} + {\bf{n}} \,.
\end{equation}
The term $M\epsilon$ describes the contributions to the residuals 
from perturbations to the timing model. 
The term $U{\bf{j}}$ describes noise that is correlated for observations made at the same time 
at different frequencies and uncorrelated over different observing epochs, 
while ${\bf{n}}$ describes uncorrelated white noise from TOA measurement uncertainties.
The term $F{\bf{a}}$ describes red noise, 
including both red noise intrinsic to the pulsar 
and a common red noise signal common to all pulsars (such as a GW signal). 
We model the red noise as a Fourier series,
\begin{equation}
	F {\bf{a}} = \sum_{j=1}^N \left[ a_j \sin\left( \frac{2\pi j t}{T} \right) + b_j \cos \left( \frac{2\pi j t}{T} \right) \right] \,,
\end{equation}
where $N$ is the number of Fourier modes used (typically $N=30$) 
and $T$ is the span of the observations.

The optimal statistic is constructed from the autocovariance and cross-covariance matrices 
$C_a$ and $S_{ab}$, 
\begin{eqnarray}
	C_a &=& \left\langle \res_a \res_a^T \right\rangle \,, \\
	S_{ab} &=& \left. \left\langle \res_a \res_b^T \right\rangle \right|_{a \neq b} \,,
\end{eqnarray}
where $\res_a$ is a vector of the residuals of the $a$th pulsar in the PTA. 
For the GWB with power spectral density (PSD) $P_\mathrm{gw}(f)$ 
and overlap reduction function (ORF) $\Gamma_{ab}$, 
the cross-covariance matrices are
\begin{equation}
	S_{ab} = F_a \, \phi_{ab}^\mathrm{gw} \, F_b^T \,,
\end{equation}
where
\begin{equation}
	\phi_{ab}^\mathrm{gw} = \Gamma_{ab} \, P_\mathrm{gw}(f) \,.
	\label{eq:phi_gw}
\end{equation}
The ORF is the HD curve \citep{hd1983}, 
\begin{eqnarray}
	\Gamma_{ab} &=& \frac{1}{2} \left[ 1 - \frac{1}{2} \left( \frac{1-\cos\theta_{ab}}{2} \right) \right. \nonumber \\
		&& \left. + 3 \left( \frac{1-\cos\theta_{ab}}{2} \right) \ln \left( \frac{1-\cos\theta_{ab}}{2} \right) \right]  \,,
\end{eqnarray}
where $\theta_{ab}$ is the angle between the pulsars. 
We model the PSD of the GWB as a power law:
\begin{equation}
	P_\mathrm{gw}(f) = \frac{\Agw^2}{12\pi^2} \left( \frac{f}{f_\mathrm{yr}} \right)^{-\gamma} \,,
\end{equation}
where $\gamma = 13/3$ assuming SMBHBs evolve solely due to GW emission and $f_\mathrm{yr} \equiv 1/(1\,\mathrm{yr})$. 
The optimal statistic $\hat{A}^2$ is given by
\begin{equation}
	\hat{A}^2 = \frac{\sum_{ab} \res_a^T {C_a^{-1} \tilde{S}_{ab} C_b^{-1} \res_b}}{\sum_{ab} \Tr \left( C_a^{-1} \tilde{S}_{ab} C_b^{-1} \tilde{S}_{ba} \right) } \,,
\end{equation}
where $\tilde{S}_{ab}$ is the amplitude-independent cross-correlation matrix,
\begin{equation}
	\Agw^2 \tilde{S}_{ab} = S_{ab} \,.
\end{equation}
This definition of the optimal statistic ensures that  
$\langle \hat{A}^2 \rangle = \Agw^2$. 
If $\Agw=0$, the variance of the optimal statistic is
\begin{equation}
	\sigma_0 = \left[ \sum_{ab} \Tr \left( C_a^{-1} \tilde{S}_{ab} C_b^{-1} \tilde{S}_{ba} \right) \right]^{-1/2} \,.
\end{equation}
For a measured value of $\hat{A}^2$, 
the significance of $\hat{A}^2 \neq 0$ is given by the signal-to-noise ratio (SNR) 
\begin{equation}
	\rho = \frac{\sum_{ab} \res_a^T {C_a^{-1} \tilde{S}_{ab} C_b^{-1} \res_b}}{ \left[ \sum_{ab} \Tr \left( C_a^{-1} \tilde{S}_{ab} C_b^{-1} \tilde{S}_{ba} \right) \right]^{1/2}} \,.
\end{equation}

When constructing the residuals $\res_a$, 
we typically fix the red noise parameters to the values that 
maximize the single-pulsar likelihood. 
However, this leads to a bias in the optimal statistic 
because the individual red noise and common red noise parameters 
are highly covariant, 
with the optimal statistic computed using fixed red noise parameters 
systematically lower than the true value of $\Agw^2$. 
In this section, we compare three techniques for computing the optimal statistic. 
First, we fix the individual pulsars' red noise parameters to the 
maximum-likelihood values from individual Bayesian pulsar noise analyses. 
Second, we fix the pulsars' red noise parameters to the 
values that jointly maximize the likelihood for a Bayesian analysis 
of all of the pulsars in our PTA 
that searches over the pulsars' red noise parameters and 
a common red process. 
For the noise-marginalized method, we draw values of the pulsars' 
red noise parameters from the posteriors generated by the common Bayesian analysis.

We use these methods to compute the optimal statistic 
for simulated ``NANOGrav-like'' data sets 
consisting of 18 MSPs with observation times, 
sky positions, and noise properties 
matching the 18 longest-observed pulsars in the 
NANOGrav 11-year data set \citep{abb+17}. 
We include white noise for all pulsars, plus 
red noise parametrized as a power law, 
\begin{equation}
	P_a(f) = \frac{A_\mathrm{red}^2}{12\pi^2} \left( \frac{f}{f_\mathrm{yr}} \right)^{-\gamma} \,,
\end{equation}
for those pulsars that show evidence of red noise 
(see Table~\ref{tab:sim} for more details).
We use the PTA data analysis package 
PAL2 
\citep{evh17a}
to perform the noise analyses and compute the optimal statistic.

\begin{table}[!tb]
	\setlength{\tabcolsep}{5pt}
	\caption{Pulsar parameters used in simulated PTA data sets.}
	\begin{center}
	\begin{tabular}{ccccc}
		\hline\hline
    		Pulsar	& $T_\mathrm{obs}$ (yrs) & $\sigma_w$ ($\mu\mathrm{s}$) & $A_\mathrm{red}$ & $\gamma_\mathrm{red}$ \\
		\hline
		J0030+0451 & 11.0 & 0.339 & -13.93 & 3.56 \\
		J0613$-$0200 & 11.0 & 0.281 & -13.14 & 1.22 \\
		J1012+5307 & 11.0 & 0.320 & -12.79 & 1.51 \\
		J1024$-$0719 & 6.0 &0.421 & $-$ & $-$ \\
		J1455$-$3330 & 11.0 & 0.773 & $-$ & $-$ \\
		J1600$-$3053 & 8.0 & 0.146 & $-$ & $-$ \\
		J1614$-$2230 & 7.0 & 0.261 & $-$ & $-$ \\
		J1640+2224 & 11.0 & 0.202 & $-$ & $-$ \\
		J1713+0747 & 11.0 & 0.093 & -14.14 & 1.58 \\
		J1741+1351 & 6.0 & 0.106 & $-$ & $-$ \\
		J1744$-$1134 & 11.0 & 0.096 & $-$ & $-$ \\
		B1855+09 & 11.0 & 0.218 & -13.75 & 3.54 \\
		J1853+1303 & 7.0 & 0.215 & $-$ & $-$ \\
		J1909$-$3744 & 11.0 & 0.034  & -13.84 & 1.74  \\
		J1918$-$0642 & 11.0 & 0.342 & $-$ & $-$ \\
		J2010$-$1323 & 7.0 & 0.413 & $-$ & $-$ \\
		J2145$-$0750 & 11.0 & 0.281 & -12.69 & 1.30 \\
		J2317+1439 & 11.0 & 0.160 & $-$ & $-$ \\
    		\hline\hline
	\end{tabular}
	\end{center}
	\label{tab:sim}
\end{table}

\begin{figure}[tb]
	\includegraphics[width=0.9\columnwidth]{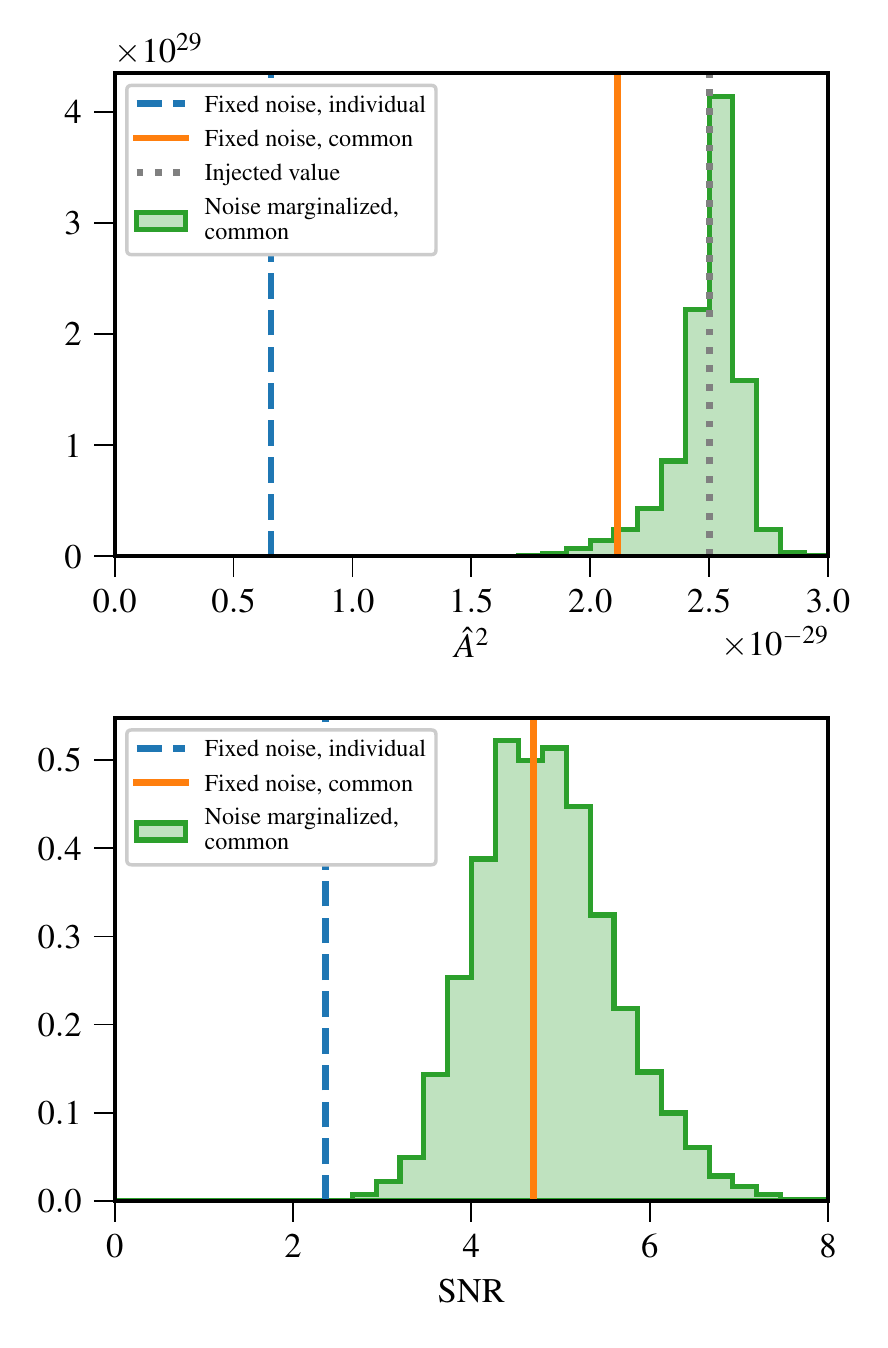}
	\caption{Optimal statistic 
			for a simulated PTA data set containing a GWB with $\Agw = 5\times10^{-15}$. 
			The fixed-noise analysis using the individual noise values (dashed blue lines) 
			systematically underestimates $\hat{A}^2$, while the 
			fixed-noise analysis using the common noise values (solid orange lines) 
			and the noise-marginalized analysis (green histograms) 
			more accurately recover $\Agw$.}
	\label{fig:os_dataset_sample}
\end{figure}

Figure~\ref{fig:os_dataset_sample} shows the fixed-noise and noise-marginalized 
optimal statistic for a simulation with a GWB with $\Agw=5\times10^{-15}$. 
For this particular realization of the GWB, 
the fixed-noise analysis 
using the individual noise results gives 
$\hat{A}^2 = 6.6\times10^{-30}$ with $\mathrm{SNR} = 2.4$, 
and the fixed-noise analysis using the common noise results gives 
$\hat{A}^2 = 2.1\times10^{-29}$ with $\mathrm{SNR} = 4.7$. 
The noise-marginalized analysis gives $\hat{A}^2 = (2.5 \pm 0.1)\times10^{-29}$ 
with $\mathrm{SNR}=4.8\pm0.8$. 
The value of $\hat{A}^2$ from the fixed-noise analysis using the individual noise results 
is significantly lower than the injected level of the GWB, while 
the values of $\hat{A}^2$ from the fixed-noise analysis using the common noise results 
and the noise-marginalized analysis are in good agreement with each other and the injected value. 
The fixed-noise analysis using the individual noise results also gives a significantly lower SNR 
than the other two. 

\begin{figure}[tb]
	\includegraphics[width=0.9\columnwidth]{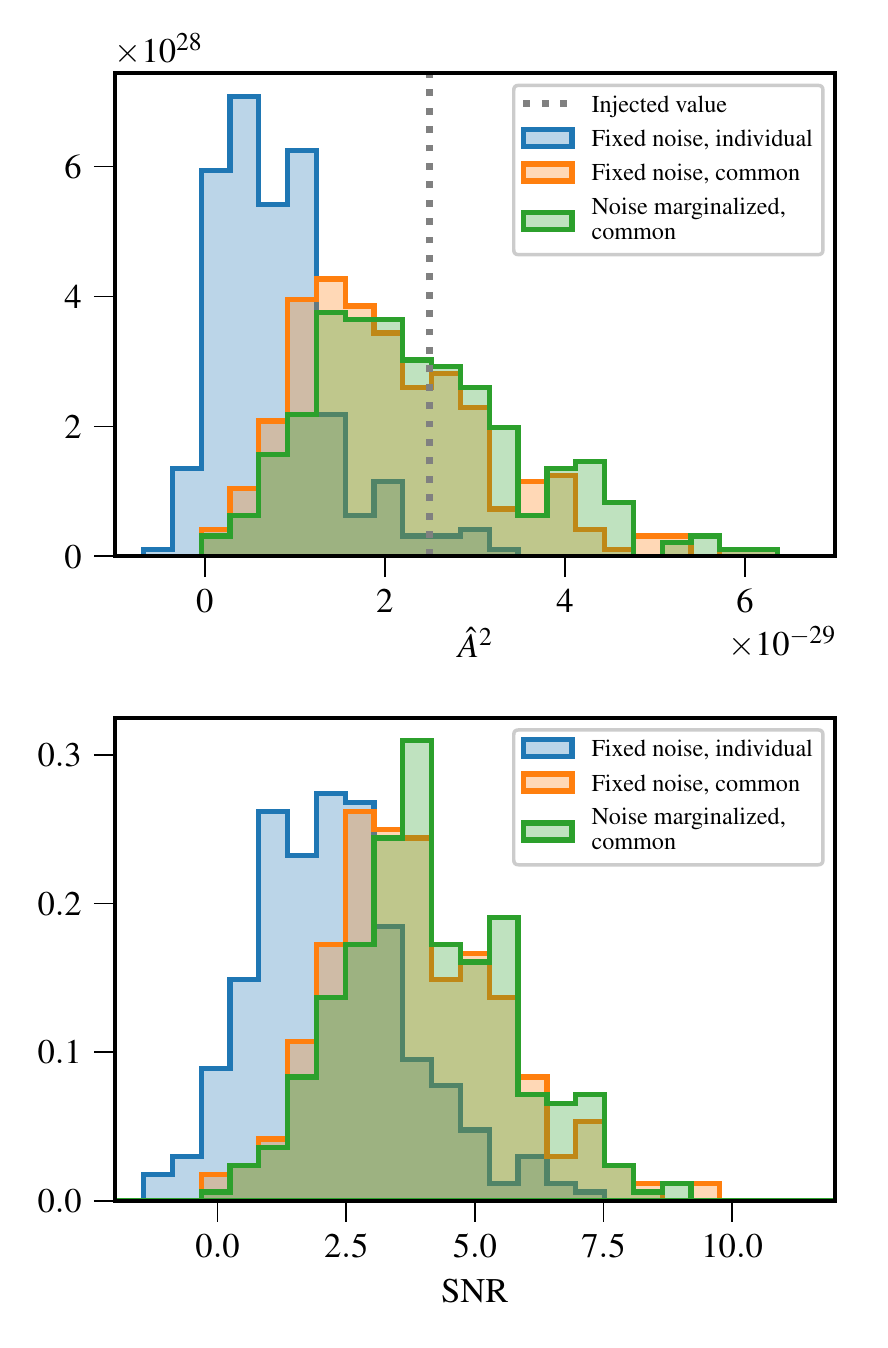}
	\caption{Optimal statistic and SNR for 300 simulated data sets 
			containing a GWB with $\Agw = 5\times10^{-15}$. 
			The fixed-noise analysis using the individual noise values (blue) 
			systematically underestimates $\hat{A}^2$, while both the 
			fixed-noise analysis using the common noise values (orange) 
			and the noise-marginalized analysis (green) 
			accurately recover $\Agw$.}
	\label{fig:os_datasetstats}
\end{figure}

In Fig.~\ref{fig:os_datasetstats} we show the optimal statistic 
for 300 different realizations of a GWB with $\Agw = 5\times10^{-15}$ 
computed using the three techniques described above. 
For the noise-marginalized analysis, we plot 
the mean values of $\hat{A}^2$ and $\rho$. 
Using the noise values from individual noise analyses 
systematically underestimates the strength of the GWB, 
while using the noise values from a common noise analysis 
more accurately recovers the injected value. 
The fixed-noise analysis using the individual noise results finds 
$\hat{A}^2 = (7.9 \pm 6.8) \times10^{-30}$ and $\rho = 2.3 \pm 1.5$, 
averaging over realizations of the GWB. 
The fixed-noise and noise-marginalized analyses 
using the common noise results both give 
$\hat{A}^2 = (2.4\pm1.2)\times10^{-29}$ and $\rho = 4.1 \pm 1.7$.

\begin{figure*}[tb]
	\includegraphics[width=0.95\textwidth]{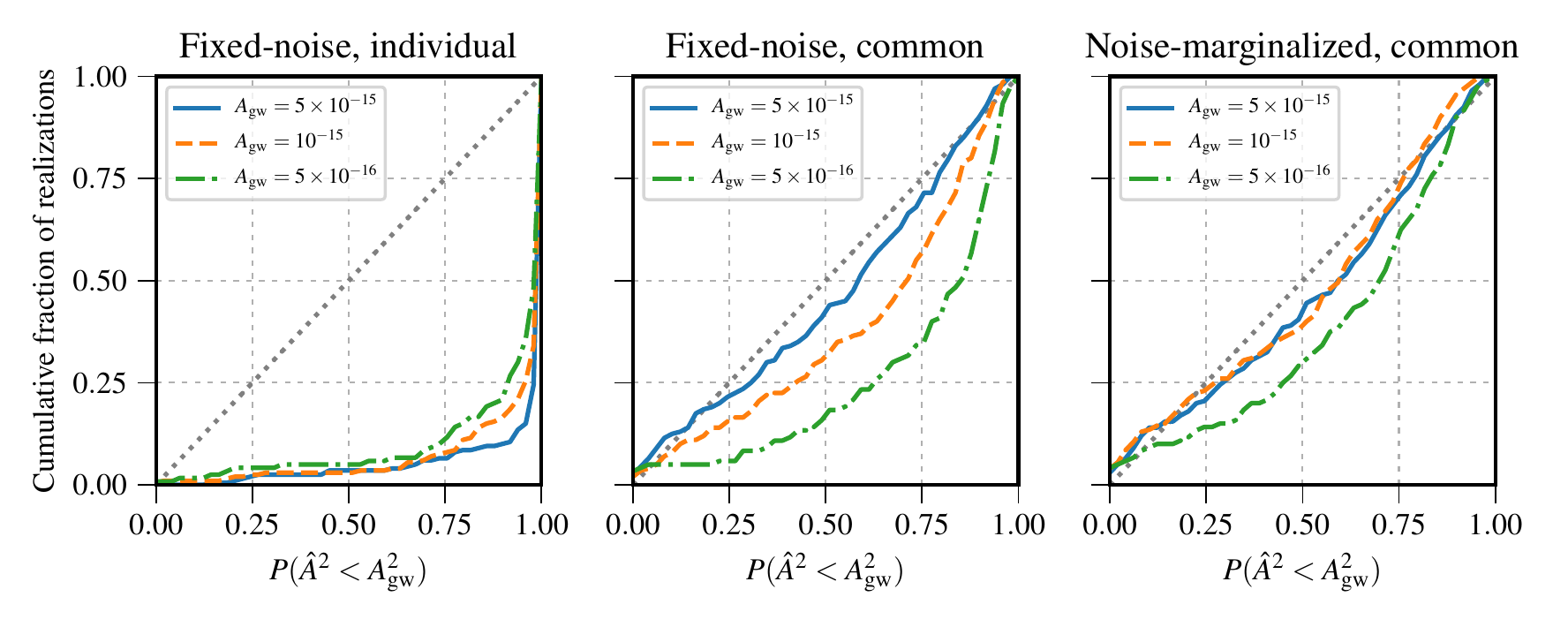}
	\caption{P--P plot showing the cumulative fraction of simulations for which $\Agw^2$ lies within 
			a given confidence interval of the measured $\hat{A}^2$. 
			The probability distribution of $\hat{A}^2$ is assumed to be a Gaussian 
			with variance $\sigma^2_{\hat{A}^2}$. 
			The fixed-noise optimal statistic using the individual and common noise results 
			both give biased values of $\hat{A}^2$, particularly for small values of \Agw, 
			while the noise-marginalized optimal statistic gives more accurate values of $\hat{A}^2$ 
			over a large range of injected values of \Agw.}
	\label{fig:os_compare}
\end{figure*}

The fixed-noise and noise-marginalized analyses 
using the common noise results give the same results 
for $\Agw = 5\times10^{-15}$, 
but there are differences between them when analyzing data sets 
containing smaller injected values of $\Agw$. 
In Fig.~\ref{fig:os_compare} we show a P--P plot 
of the cumulative fraction of simulations for which the injected $\Agw^2$ lies within 
a given confidence interval of the measured $\hat{A}^2$. 
The confidence interval of $\hat{A}^2$ is determined 
assuming $\hat{A}^2$ follows a Gaussian distribution, 
with mean and variance $\sigma^2_{\hat{A}^2}$ 
taken from the distribution for $\hat{A}^2$ 
found from our 300 realizations of the GWB (i.e.,~the top panel of Fig.~\ref{fig:os_datasetstats}).
If $\hat{A}^2$ has a Gaussian distribution centered around $\Agw^2$, 
the curves should lie along a straight line with slope equal to unity (the dotted, diagonal lines in Fig.~\ref{fig:os_compare}).

We compare the three methods for computing the optimal statistic for 
simulations with $\Agw=5\times10^{-15}$, $\Agw=10^{-15}$, and $\Agw=5\times10^{-16}$. 
The fixed-noise optimal statistic using the individual noise results systematically underestimates $\hat{A}^2$ 
(Fig.~\ref{fig:os_compare}, left panel). 
The fixed-noise optimal statistic using the common noise results 
recovers $\hat{A}^2$ well for large values of $\Agw$, but for small values it also underestimates $\hat{A}^2$ 
(Fig.~\ref{fig:os_compare}, middle panel). 
The noise-marginalized optimal statistic provides 
the most accurate estimate of $\hat{A}^2$ 
over the range of \Agw\ considered here, 
although it still slightly underestimates \Agw\ 
(Fig.~\ref{fig:os_compare}, right panel).

\section{Monopole and Dipole Spatial Correlations}
\label{sec:spatial}

The optimal statistic is particularly well-suited 
to compare multiple spatial correlation relations 
because using a different spatial correlation only requires changing the ORF 
in Eq.~\eqref{eq:phi_gw}. 
\citet{thk+2016} demonstrated how the optimal statistic can be altered to fit for 
multiple spatial correlations at once in order to mitigate common noise sources such as 
clock error and ephemeris error. 
Here we take a different approach -- rather than simultaneously fitting 
for signals with different spatial correlations, 
we look at how well we can distinguish between different spatial correlations 
by computing the optimal statistic with monopole and dipole spatial correlations 
for the same simulated data sets as in the previous section. 
For a monopole signal, the ORF becomes simply
$\Gamma_{ab} = 1$, 
while for a dipole signal, the ORF becomes
$\Gamma_{ab} = \cos\theta_{ab}$.

Our ability to distinguish between different spatial correlations 
depends on the strength of the GWB 
and the angular separations between pulsar pairs, $\theta_{ab}$. 
We can determine the overlap between ORFs corresponding to different spatial correlations 
by computing the ``match statistic'' \citep{cs2016},
\begin{equation}
	\bar{M} = \frac{\sum_{a,b \neq a} \Gamma_{ab} \Gamma'_{ab}}{\sqrt{ \left( \sum_{a, b \neq a} \Gamma_{ab} \Gamma_{ab} \right) \left( \sum_{a, b \neq a} \Gamma'_{ab} \Gamma'_{ab} \right)}} \,,
\end{equation}
where $\Gamma$ and $\Gamma'$ are two different ORFs. 
For the 18 pulsars used in these simulations, the 
match statistic for monopole and HD correlations is $\bar{M} = 0.264$, 
and the match statistic for dipole and HD correlations is $\bar{M} = 0.337$. 
These match statistics describe a fundamental limit on our ability 
to identify the spatial correlations of a common red signal as HD 
rather than monopole or dipole 
that depends only on the number of pulsars in our PTA and their sky positions.

\begin{figure}[t]
	\includegraphics[width=0.9\columnwidth]{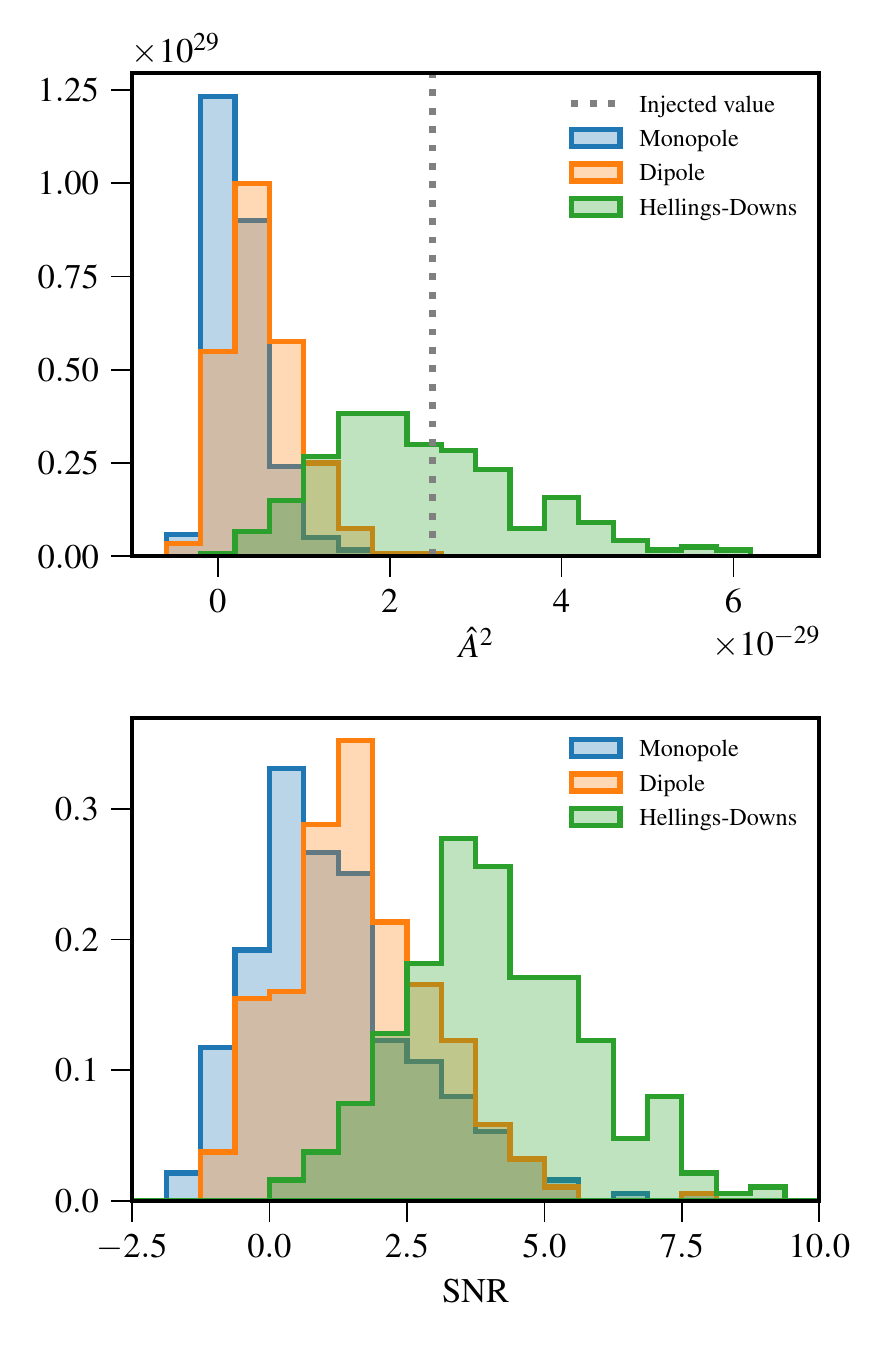}
	\caption{Noise-marginalized mean optimal statistic and mean SNR for 300 simulated data sets 
			containing an injected GWB with $\Agw = 5\times10^{-15}$. 
			We compare the values of the mean optimal statistic and the SNR found 
			using monopole (blue), dipole (orange), and HD (green) spatial correlations. 
			The dashed vertical line indicates the injected value, $\hat{A}^2 = 2.5 \times 10^{-29}$.}
	\label{fig:os_ORF}
\end{figure}

Figure~\ref{fig:os_ORF} shows the noise-marginalized 
mean value of $\hat{A}^2$ and the mean SNR 
computed assuming monopole, dipole, and HD spatial correlations 
for 300 simulated data sets. Using a monopole or dipole ORF 
gives a lower value for the mean optimal statistic and mean SNR compared to the 
HD ORF. 
Using HD spatial correlations gives $\hat{A}^2 = (2.4 \pm 1.1) \times 10^{-29}$, 
while using monopole spatial correlations gives $\hat{A}^2 = (2.5 \pm 3.2) \times 10^{-30}$, 
and dipole spatial correlations gives $\hat{A}^2 = (5.2 \pm 4.4) \times 10^{-30}$. 
We find a noise-marginalized mean SNR above 1.0 in 97\% of our simulated data sets 
using the HD ORF, and in 50\% and 68\% of our simulated data sets 
using the monopole and dipole ORFs, respectively. 
The mean SNR using the HD ORF, averaged over realizations of the GWB, is 4.1, 
and we find an SNR greater than this using the monopole and dipole ORFs in just 
3\% and 3.5\% of our simulations, respectively.

\begin{figure}[t]
	\includegraphics[width=0.9\columnwidth]{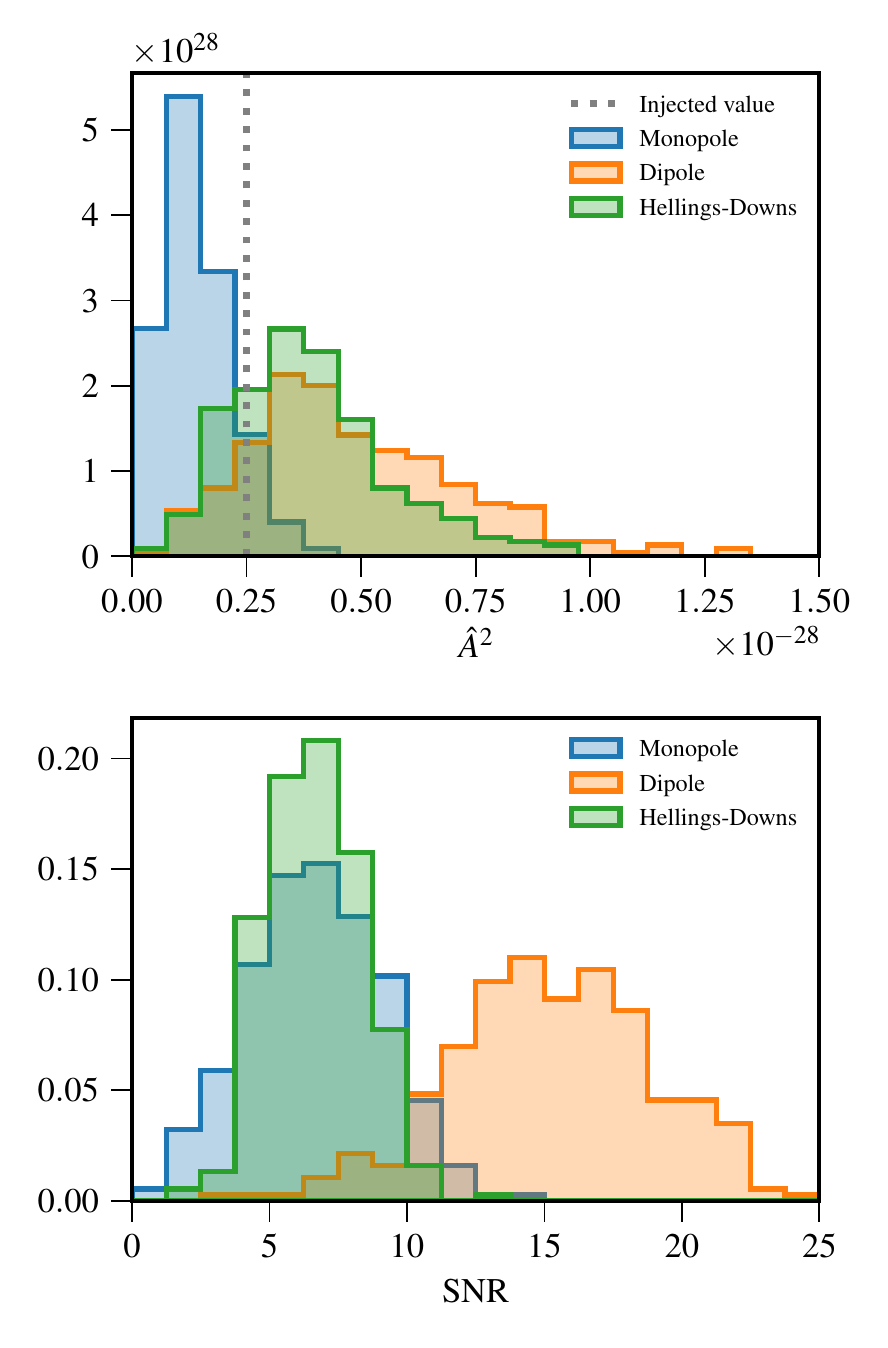}
	\caption{Noise-marginalized mean optimal statistic and mean SNR for 300 simulated data sets 
			containing an injected stochastic signal with dipole spatial correlations and $A = 5\times10^{-15}$. 
			We compare the values of the optimal statistic and mean SNR 
			found using monopole (blue), dipole (orange), 
			and HD (green) spatial correlations. 
			The dashed vertical line indicates the injected value, $\hat{A}^2 = 2.5 \times 10^{-29}$.}
	\label{fig:os_dipole}
\end{figure}

This overlap between the monopole, dipole, and HD ORFs also means that 
a common red process that does not have HD correlations  
may be confused for a GWB. Figure~\ref{fig:os_dipole} shows the results of 300 simulations 
containing a stochastic signal with dipole spatial correlations. 
Although a dipole signal has been injected, 
the HD ORF gives a mean SNR greater than 5 in 82\% of the simulations. 
However, both the monopole and HD ORFs give smaller values of 
the mean $\hat{A}^2$ and mean SNR compared to the dipole ORF. 
Furthermore, there are no simulations for which the mean SNR with HD ORF 
is greater than the mean SNR with dipole ORF. 
This demonstrates the importance of comparing the SNR from different spatial correlations 
when determining the type of spatial correlations present.

\section{Sky Scrambles}
\label{sec:skyscrambles}

The significance of spatial correlations can also be tested with ``sky scrambles,'' 
where the ORF is altered in order to simulate changing the pulsars' positions \citep{cs2016,tlb+2017}. 
The scrambled ORFs are required to have small values of $\bar{M}$ 
with the true ORF and each other so that they form a nearly orthogonal set. 
This ensures that the distribution of $\hat{A}^2$ computed using the scrambled ORFs 
forms the null hypothesis distribution. 
\citet{tlb+2017} showed how sky scrambles affect the Bayes' factor for simulated data sets. 
We performed a similar analysis using frequentist methods.

We generated 725 scrambled ORFs 
using a Monte Carlo algorithm. 
We required the scrambled ORFs to have $\bar{M} < 0.2$ 
with respect to the true ORF and each other. 
This threshold was chosen to be comparable to the match statistics 
between the HD ORF with monopole and dipole ORFs 
given in Sec.~\ref{sec:spatial}. 
We did not choose a smaller threshold because significantly more time 
would have been needed to generate 725 scrambled ORFs. 
For each simulation, we computed the 
noise-marginalized mean optimal statistic and mean SNR 
for each scrambled ORF, 
and compared them to the values found using the true ORF.

\begin{figure}[t]
	\includegraphics[width=0.9\columnwidth]{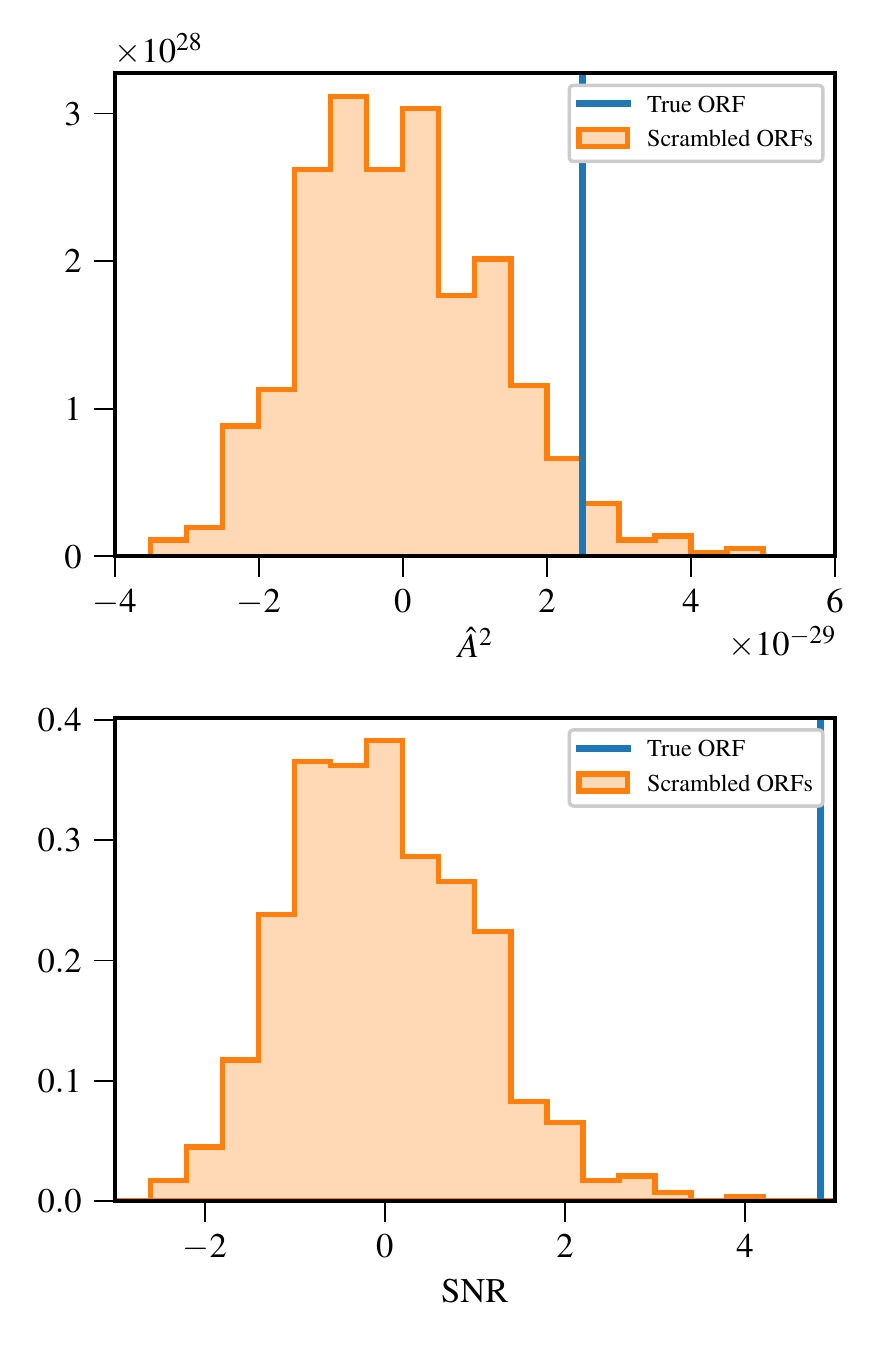}
	\caption{Comparison between the noise-marginalized mean optimal statistic and mean SNR 
			with and without sky scrambles for a simulated data set 
			containing a GWB with $\Agw = 5\times10^{-15}$. 
			None of the 725 scrambled ORFs gave a mean SNR as large 
			as the mean SNR using the true ORF ($p < 0.0014$).}
	\label{fig:skyscrambles_dataset_sample}
\end{figure}

\begin{figure}[tb]
	\includegraphics[width=0.9\columnwidth]{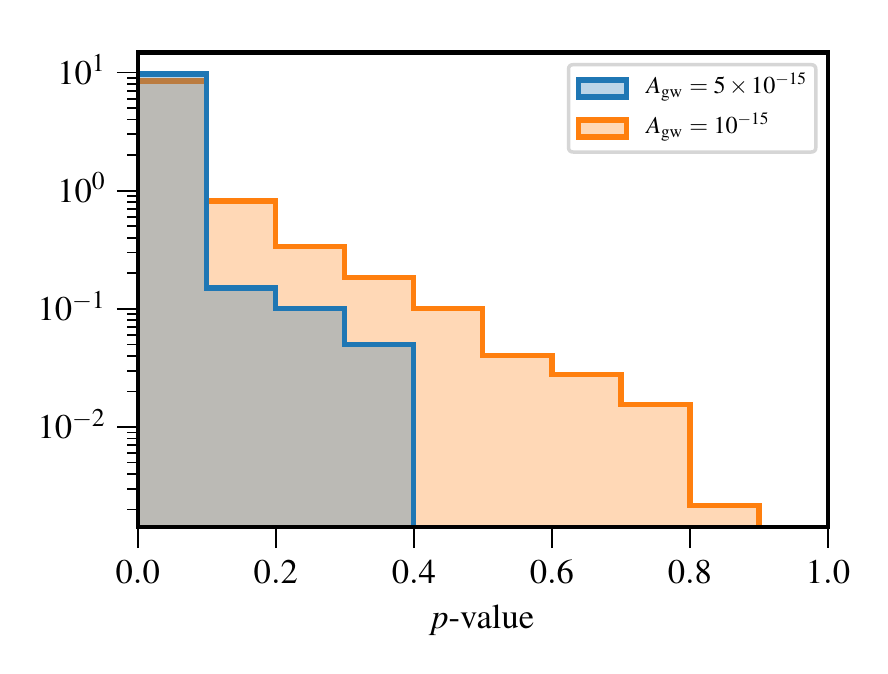}
	\caption{Distribution of $p$-values for the noise-marginalized optimal statistic mean SNR 
			using the true ORF compared to 725 sky scrambles from 300 realizations of the GWB. 
			We show results for simulations with $\Agw=5\times10^{-15}$ (blue) and $\Agw=10^{-15}$ (orange).}
	\label{fig:skyscrambles_pvalues}
\end{figure}

Figure~\ref{fig:skyscrambles_dataset_sample} shows the results of a sky scramble analysis 
for a sample data set with $\Agw=5\times10^{-15}$. 
For this particular realization of the GWB, 
none of the 725 scrambled ORFs resulted in a mean SNR greater than 
the mean SNR using the true ORF ($p < 0.0014$).
In Fig.~\ref{fig:skyscrambles_pvalues}, we plot the distribution of $p$-values of the 725 sky scrambles 
for 300 realizations of the GWB. 
For a GWB with $\Agw = 5\times10^{-15}$, 95\% of the simulations have $p \leq 0.05$, 
and 74\% of the simulations have $p \leq 0.003$. 
For a GWB with $\Agw = 10^{-15}$, 76\% of the simulations have $p \leq 0.05$, 
and 39\% have $p \leq 0.003$. 
This shows that for smaller values of \Agw, 
there is a greater chance that noise fluctuations will appear 
to have the spatial correlations of the GWB.

\section{Conclusion}
\label{sec:conclusion}

The definitive signature of a GWB in PTA data is spatial correlations 
described by the HD curve. 
Searching for these using a full Bayesian approach is computationally expensive, 
requiring many weeks on a supercomputing cluster. 
In contrast, the optimal statistic can be computed in seconds. 
In this paper, we introduce an improved method 
for computing the optimal statistic, 
which uses the output from a Bayesian analysis 
for individual and common red signals 
to marginalize the optimal statistic over the individual pulsars' red noises. 
As shown in Sec.~\ref{sec:marg_os}, 
the noise-marginalized optimal statistic more accurately recovers the GWB amplitude 
than the fixed-noise optimal statistic, which underestimates the GWB amplitude 
when significant red noise is present in some pulsars.

Although the noise-marginalized optimal statistic 
requires computing the optimal statistic thousands of times, 
it is still many orders of magnitude faster than a Bayesian search. 
Furthermore, the results from a single Bayesian analysis, 
which are needed to marginalize over the red noise parameters, 
can be used to compute the optimal statistic for many different spatial correlations. 
In Sec.~\ref{sec:spatial} we use the noise-marginalized optimal statistic 
to compare the strength of monopole, dipole, and HD correlations 
in simulated PTA data with a GWB. 
In Sec.~\ref{sec:skyscrambles} we use the noise-marginalized optimal statistic 
to perform sky scramble analyses, where we compare the mean SNR 
computed using the true ORF to the mean SNR computed using scrambled ORFs 
and measure the significance of HD spatial correlations through the $p$-value.

The primary strength of the optimal statistic is how quickly it can be computed. 
This is useful for analyses where the significances of many spatial correlations are compared, 
as with the sky scrambles. 
An upcoming paper will use the noise-marginalized optimal statistic 
to determine how well the spatial correlations corresponding to alternate GW polarizations 
can be measured. 
It also makes the optimal statistic a valuable tool for analyzing simulations 
where many realizations of the GWB are compared. 
The noise marginalization technique described in this paper is key 
to being able to accurately measure the GWB with the optimal statistic 
for real PTAs and realistic PTA simulations, 
for which red noise is significant.

\acknowledgments
We thank Joe Lazio, Andrea Lommen, Joe Romano, Xavier Siemens, 
and Jolien Creighton for useful discussions. 
This work was supported by NSF Physics Frontier Center Grant 1430284. 
JAE was partially supported by NASA through Einstein Fellowship Grant PF4-150120. 
We are grateful for computational resources provided by the Leonard E.~Parker
Center for Gravitation, Cosmology and Astrophysics at the University of
Wisconsin-Milwaukee, which is supported by NSF Grants 0923409 and 1626190.

\bibliographystyle{apsrev}
\bibliography{master}

\end{document}